\documentstyle[aps,preprint]{revtex}

\begin{document}

\title{In search for natural wormholes}
\author{Luis A. Anchordoqui$^{1}$\thanks{doqui@venus.fisica.unlp.edu.ar}, 
Gustavo E. Romero$^2$\thanks{Member of
CONICET}, Diego F.
Torres$^1$ and I. Andruchow$^3$}

\address{$^{1}$Departamento de F\'{\i}sica, Universidad Nacional de La Plata, C.C.
67, 1900 La Plata, Argentina}
\address{$^{2}$Instituto Argentino de Radioastronom\'{\i}a, C.C. 5, 1894 Villa
Elisa, Argentina}
\address{$^3$Facultad de Ciencias Astron\'omicas y Geof\'{\i}sicas, Universidad Nacional
de La Plata, Paseo del Bosque S/N, 1900 La Plata, Argentina}

\maketitle
\begin{abstract}
We have investigated 631 time profiles of gamma ray bursts from the BATSE
database searching for observable signatures produced by microlensing
events related to natural wormholes. The results of this first search of
topologically nontrivial objects in the Universe can be used to constrain
their number and mass. 

\noindent {\it PACS number(s):} 98.62.Sb, 04.20.Gz
\end{abstract}

\newpage

\section{Introduction}

Wormholes are nontrivial topological configurations of spacetime that can
be represented by solutions of Einstein field equations with stress-energy
tensor fields that somewhere violate the so-called average null energy
condition (see Ref. \cite{visser-book} for a detailed discussion).
Although microscopic violations of the energy conditions are well known
(e.g. the Casimir effect), it is far from clear whether stable,
macroscopic wormholes can naturally exist in the Universe. 
One of the ways in which one may obtain violations to the energy conditions
is via a scalar fields coupled to gravity (see for instance \cite{NOS1} and
references therein).

Wormhole formation at a late cosmic time requires Lorentzian topology
change in space, something that appears to be more than problematic to
most physicists because it implies causality violations
\cite{geroch,hawking}. However, if wormholes are created altogether with
spacetime and not formed by astrophysical processes, one could expect a
cosmological population of these objects without the uncomfortable
predictions of topology change theorems.

In a couple of recent papers we have discussed the observable effects that
could arise from an intergalactic population of natural wormholes
\cite{nosotros1,nosotros2}. Since wormhole's mouths could have a total
negative mass, they should exert a repulsive gravitational force that can
provide very peculiar microlensing events when acting upon the light of
compact, background sources \cite{cramer}. Extragalactic wormholes with
absolute masses of $\sim$ 1 M$_{\odot}$ would produce very compact
Einstein rings, in such a way that just small, ultraluminous sources like
the $\gamma$-ray emitting core of quasars (typical size $10^{14}-10^{15}$ cm)
might result gravitationally magnified. We have shown in Ref. \cite{nosotros2} that
the lightcurve signature of wormhole microlensing events of this sort very much
resembles some kinds of gamma ray bursts (GRBs).

When a negative mass lens crosses the line of sight to a distant quasar,
dragging the caustic pattern along with it, two bursting $\gamma$-ray
events will appear in the observer's frame: the first one is the specular
image of a fast-rise-exponential-decay (FRED) burst, whereas the second,
after a period
of stillness that can last several years, is a pure FRED event. In our
previous study \cite{nosotros1}, we have used the
available database of GRB observations gathered by the BATSE instrument,
part of the
Compton satellite, to set an upper limit to the amount of
negative mass (under the form of compact objects of astrophysical size) in
the Universe. Such limit results as low as $|\rho|\leq 2\times10^{-33}$ g
cm$^{-3}$ with the most optimistic assumptions.    

In the present paper we give a step further and embark on the first
detailed search for individual wormhole signatures in astronomical
databases. GRBs produced by natural wormholes can be differentiated from
those originated in fireballs because of two very definite properties: 1)
they repeat, and 2) one of the repeating bursts has an anti-FRED time
profile, something that cannot be the result of an explosive event
\cite{romero} (the companion burst must display a FRED-like lightcurve).

We have quantitatively analysed a subsample of the GRBs included in the
BATSE 3B cataloge with the aim of identify events that could be
unequivocally attributed to wormhole lensing. In what follows we present
the results we have obtained.   

\section{Data analysis}

We have analysed a sample of 631 bursts from BATSE 3B catalog whose global
symmetry properties were already discussed by Link \& Epstein \cite{link},
and Romero et al. \cite{romero}. This sample contains both faint and
bright bursts, spanning 200-fold range in peak flux. PREB + DISC data
tapes at 64 ms time resolution, with four energy channels, were used in
the analysis.

Since the variety of burst profiles is huge and simple visual inspection
can be misleading, we have used the skewness function $\cal{A}$ introduced
by Link \& Epstein \cite{link} in order to separate those GRBs with
anti-FRED profiles. The skewness is basically defined as the third moment
of the individual burst time profile and can be directly computed from the
observational data as in Ref. \cite{link}. 
Negative values of $\cal{A}$ correspond to events with
slower rising than decaying timescales, thus showing a peculiar asymmetric burst
(PAB).

In a first step, we estimated $\cal{A}$ for all GRBs in the sample at
different background cutoff levels. Just 91 out of 631 bursts present
$\cal{A}<$0 at any background. As discussed in Ref. \cite{romero}, most of
these events can be explained within the standard fireball model of GRBs
\cite{rees}. Just anti-FRED--like single peaked events remain inconsistent
with the explosive hypothesis. There are 26 of these GRBs in our sample (4.1
\%). 

Since wormhole lensing not just provides bursts with $\cal{A}<$0 but also a
repetition with specular signature, we have searched for time-space
clustering in the sample. We have found that 15 out of 26 candidates
(about 60 \%) present companions within error boxes at less than $4^{o}$
(the average positional uncertainty in BATSE catalog). We have estimated
the statistical 
significance of this level of positional coincidences through numerical
simulations of random sets of 26 events against a background
distribution of 605 GRBs. After 1500 simulations we established that the
chance associations expected in the subsample are $13.3\pm2.5$, i.e. there
is no need to claim for repetition to explain the observed coincidences at
error boxes of $4^o$. However, if positional coincidences separated by
less than $1^o$ are considered, we find that 3 out of 26 events present
companions. According to a new set of simulations, these results can be
attributed to chance only at a $2\sigma$ level. Despite the sample is too
scarce to draw any conclusion, it is worth mentioning that when a similar
study is carried out with the 91 bursts with $\cal{A}<$0 it is found that
there are just 4 positional coincidences at less than $1^o$, at $1\sigma$
confidence level. This could imply that the apparent excess is exclusively
associatted to single peaked events, as expected from the microlensing model.

In order to detect whether there is some suitable wormhole candidate
behind the above mentioned statistical analyses, we turned to the
individual single peaked events with $\cal{A}<$0 in a finer search.

\section{Results}

In Table 1 we list by trigger number all single peaked bursts with
$\cal{A}<$0 in our sample. In column 2 we indicate the trigger number of
any companion burst in the entire BATSE database within a circle of $4^o$
in radius. Columns 3 and 4 display the temporal and angular distances of
pairs of events. A negative value of $\Delta T$ means that the anti-FRED
event followed to its companion; such bursts can be eliminated as wormhole
candidates, at least over the timescales under consideration here. The
final column in the table lists the sign of $\cal{A}$, when defined, for
companion bursts that belong to our subsample. Bursts with no entries in
this column where not analyzed in the present study, circunscribed to the
previously defined set of 631 GRBs.

We shall consider as candidates for wormhole microlensing just events
with companions that present $\cal{A}>$0 at all levels of background
(notice that this is a very restrictive criterion, and  
eliminates the event mentioned in
Ref. \cite{nosotros2}). This left us with only 4 candidates: \#254, \#444,
\#1924, and \#2201. A further step can be made now by detecting
active galactic nuclei (AGNs) within the error boxes of the bursts. These
AGNs would constitute the potential background sources of gamma rays.  

In Table 2 we list pairs of GRBs along with the AGNs (namely compact QSOs) within
the BATSE field. We also indicate the morphological type of the bursts
with $\cal{A}>$0. As can be appreciated from this table, three pairs of
events present quasars in their fields: \#254, \#444, and \#1924. These
are the stronger candidates for wormhole microlensing events in our sample
of 631 bursts. None of them, however, can be considered as a certain
identification because the profiles of the second bursts in each pair are
not exact FREDs despite presenting $\cal{A}>$0 at all levels. These
bursts have profiles with some substructure which is not present in the
first event of the pair. Although such a fine substructure could be an
effect of the different light propagation paths (the light can be
exposed during its travel to lensing effects by ordinary matter that might
result in a 
distortion of the original profile \cite{prop}) or even an artifact due to
the
different orientation of the spacecraft at the detection times, we think
that the evidence is not strong enough to claim for an indisputable
identification. 

In the case of the pair \{\#2201, \#2679\}, both bursts are single peaked
and present the correct symmetry in their profiles. By other hand, there
are no cataloged AGNs in the corresponding sky field. This would not be an
insurmountable problem for wormhole microlensing because even very weak
and normally undetected QSOs can be enhanced by caustic crossing in such a
way as to appear as a bright source during a few seconds \cite{nosotros2}.
However, in this particular case, the flux ratio of both bursts
(which should be similar as it comes from the same source)
is too far
from unity as to make a case for the lensing argument.

The best candidate in the whole sample is the pair \{\#254, \#2477\}.
They present correct symmetry properties and unity flux ratio. The event
\#2477 has a substructure that make of it not a perfect FRED, but this, as
it was mentioned, could stem from propagation effects. Two AGNs, with
redshifts of 0.15 and 1.52, are present within the positional 
uncertainties.\footnote{It should also be pointed out the FRED--anti-FRED pair,  
\#688 and \#2788, which is located
at the same position in the sky (within 4$^o$) together with 
156 QSOs. These two bursts might be produced by two different microlensing
phenomena with timescales that span out the BATSE operation time.}   

The greatest difficulty at present time is the huge positional uncertainty,
something that will be significantly improved in ten years time.
The results of our search, although not conclusive, are sufficiently
suggestive as to encourage new studies over larger samples and with the more
accurate detectors of the forthcoming generation of gamma ray satellites.

\section{Final comment}

Our search for natural wormholes through microlensing was sensitive to
timescales up to 3.5 years. Repetition of anti-FRED events over longer
scales can not be ruled out. Tegmark et al. \cite{teg} have made a
repetition study on the entire BATSE sample concluding that a repetition
level of $\sim5$ \% with timescales of a few years is compatible with the
current data. Our results show that, if repetition is associated to
wormhole microlensing alone, it could reach, at most, a level $\sim 4\%$
over timescales larger than 3.5 years. At shorter scales, wormhole-induced
repetitions are constrained at a level $<0.2$ \% (assuming \{\#254,
\#2477\} as the sole possible candidate).

Since microlensing timescale increases with larger masses
of the lenses, 
the absence of clear detections in our search might be
saying that wormholes, if there exist at all, have a mass distribution
peaking far beyond the few tenths of solar masses required to produce
typical microlensing events with timescales of a few years. 
If we recall
that a negative mass of the size of Jupiter 
is necessary to keep open a wormhole throat
of about 2 m in diameter \cite{visser-book}, this result could be kindly
greeted by optimistic interstellar-travel afficionados looking for larger
spacetime tunnels.

\section*{Acknowledgements}

We acknowledge Bennett Link for earlier discussions on burst temporal profiles. 
The present research has made use of the NASA/IPAC Extragalactic Database,
which is operated by the Jet Propulsion Laboratory, California Institute
of Technology, under contract with the National Aeronautics and Space
Administration. This work has been partially supported by Argentine
agencies CONICET (G.E.R., D.F.T.) (also, through
research grant PIP 0430/98 and ANPCT, G.E.R.). L.A.A. thanks FOMEC
Program for additional support.

\newpage

\begin{table}
\label{18}
\centering
\caption{Single peaked anti-FRED--like GRBs (A-GRB) and companion bursts within
the positional error boxes. Individual events are designated by BATSE
trigger numbers. $\Delta T$ is the temporal distance in
days between the two events, whereas $\Delta d$ [$^o$] is the angular
separation. The last column lists the sign of the skewness function for
four different background cutoff levels. A $\mp$ stands for bursts whose
values of ${\cal A}$
are negative but the error includes positive regions, similarly,
a $\pm$ sign means a positive skewness reaching negative values for some
regions of the error box. A little $\dag$ means that ${\cal A}$ is not
defined at the corresponding cutoff.}
\hfill
\begin{tabular}{llclc}
A-GRB & Other triggers &   $\Delta T$ & 
$\Delta d$ [$^o$] & ${\cal A}$ \\
\hline
\# 179  & & & &  \\
\# 254 & \# 1742 &  430 & 2.6  & $\pm$ $\pm$ $-$ $\mp$\\
       & \# 3113 &  1155 & 3.2  & \\
       & \# 2477 & 790 & 3.8 & + + + $\pm$ \\
\# 353 & \# 2694  & 916 & 1.1 &  \\
\# 444 & \# 2408  & 727 & 1.0 &  + + + +\\
\# 551  & \# 2431 & 738 & 2.4 &  + $-$ $-$ $-$ \\
\# 752 &   &  &  &  \\
\# 906  & & & &  \\
\# 1142 & \# 2614 & 700 & 2.7 &  + $-$ + $-$ \\
\# 1154 & \# 2124 &  397 & 2.9 &   \\
\# 1359 & & & &  \\
\# 1461\footnote{Note that NED also cite GRB 790329 within the error 
box (2.6$^o$) of trigger \# 1461.} 
& \# 1663 &  111 & 0.8 & $-$ $-$ $-$ $-$ \\
\# 1851 & \# 2319 &  224 & 3.6 &  \\
\# 1924 & \# 2830  & 522 & 0.9 &   + + $\pm$ + \\
\# 1968 & \# 2498 &  324 & 3.4 &  \\
        & \# 1480 & -201 & 3.5 &  $\dag\,$ $\,\dag\,$  $-$ + \\
\# 2161 &  &  &  & \\
\# 2163 &  &   &   & \\
\# 2201 & \# 2679 &  289 & 2.9 & + + + + \\ 
\# 2220 & \# 1975 &  -153 & 2.2 &  \\
\# 2434 &  &   &  \\
\# 2788 & \# 686  & -904 & 1.2  & + + + + \\
        & \# 1676 & -530 & 2.6 & $\mp$ + + $\pm$ \\
\# 2795 & \# 223  & -254 & 3.5 &  $\mp$ $\pm$ $\,\dag\,$ $\,\dag$ \\
     & \# 2081 &  -426 & 4.0 &  $\pm$ $\mp$ $-$ $-$\\
\# 2823 & \# 2244 & -336 & 2.6 &  $\mp$ $\pm$ $\,\dag\,$ $\,\dag$\\
\# 2846 & \# 2603 & -122 & 0.8  & + + $\pm$ $-$ \\
        & \# 2927 &  48 & 1.1 &  + + $\mp$ $\,\dag$ \\
        & \# 3132 & 179 & 2.6 & \\
\# 2918 & \# 2509 & -220 & 2.8 &  + + $-$ + \\
\# 2978 & & & &  \\
\# 2995 & \# 2945 & -33 & 1.4 &  + + $\pm$ $\pm$ \\ 
           & \# 2347 &  -373 & 3.0 &  \\
           & \# 871  & -964 & 3.1 &  + $\pm$ $\pm$ $\pm$ \\
           & \# 2394 &  -347 & 3.9 & \\
\end{tabular}
\end{table}
\begin{table}
\label{19}
\centering
\caption{Possible candidates for microlensing by wormholes
together with the corresponding background sources and the morphology
of the counterpart.
S stands for a single
peaked temporal profile whereas $S_c$ is a single peaked with a complex
substructure.}
\hfill
\begin{tabular}{llll}
Triggers and morphology & Object name & $(l,b)$ &   $z$ \\
\hline
\{\#254,\#2477\} S$_c$ & MG4 J190945+4833 & (79.3,17.1)  & 0.513 \\
                   &   MG4 J192325+4754 & (79.6,14.8)  & 1.52 \\
\{\#444,\#2408\} S$_c$ & PMN J0846+0642 & (220.4,28.7)  &  \\
                 & [HB89] 0846+100 & (217.6,30.7)  & 0.366\\
                 & [HB89] 0847+100 & (217.6,30.7)  & 2.8\\
                 & RX J0842.1+0759 & (218.5,28.2)  & \\
\{\#1924,\#2830\}  S$_c$ & 87GB 234437.2+512530 & (112.9,-9.9)  & 0.044 \\
\{\#2201,\#2679\} S &   & & \\

\end{tabular}
\end{table}

\end{document}